\documentclass[12pt]{iopart}

\usepackage{graphicx}
\usepackage{graphics}
\usepackage{grffile} 
\usepackage{epsfig}
\usepackage[leftcaption]{sidecap}

\newcommand{\vnd} {\ensuremath{V_{n\Delta}}{ }}
\newcommand{\ptt} {\ensuremath{p_{T}^{t}}{ }}
\newcommand{\pta} {\ensuremath{p_{T}^{a}}{ }}
\newcommand{\df}{\ensuremath{\Delta\phi}{ }}
\newcommand{\de}{\ensuremath{\Delta\eta}{ }}
\newcommand{\CF}{\ensuremath{C \left(\Delta \phi, \Delta \eta \right)}{ }}

\newcommand{\iaa} {\ensuremath{I_{AA}}{ }}
\newcommand{\icp} {\ensuremath{I_{CP}}{ }}
\newcommand{\raa} {\ensuremath{R_{AA}}{ }}

\renewcommand{\pt}           {\ensuremath{p_{T}{ }}}

\newcommand{\snn}          {\ensuremath{\sqrt{s_{\rm NN}}}}

\begin{document}

\title[Triggered di-hadron correlations in Pb--Pb collisions from the ALICE experiment]{Triggered di-hadron correlations in Pb--Pb collisions from the ALICE experiment}
\author{Andrew M. Adare for the ALICE collaboration}

\address{Yale University, New Haven, CT 06520}
\ead{andrew.adare@yale.edu}
\begin{abstract}
  Angular correlations between unidentified hadron trigger and
  associated particles are measured by the ALICE experiment for $0.5 <
  p_T^{t,a} < 15$ GeV, where $\ptt \geq \pta$. The pair correlation
  shapes are examined in a variety of centrality categories for pairs
  in $|\eta| < 1.0$ where $|\eta^t - \eta^a| > 0.8$. A series of
  two-particle Fourier components $V_{n\Delta} \equiv \langle
  \cos(n\Delta\phi)\rangle$ are extracted from the long-range
  azimuthal correlation functions. The sum of $n<6$ terms match the
  data. For each $n$, a fit is applied over all $p_{T}$ bins
  simultaneously to test the collectivity hypothesis $V_{n\Delta}
  \simeq v_n^t \, v_n^a$. The factorization holds at $p_{T}^{t,a}$
  below approximately 4 GeV but breaks progressively at higher
  momenta. The divergence between the data and the global fit
  quantifies the onset of nonflow dominance in long-range correlations
  due to the away side jet. The $v_n$ values from the global fit are
  in close agreement with results from more established methods. At
  higher \pt~where jet correlations dominate, the modification of
  conditional yields in central Pb--Pb collisions is measured with
  respect to $pp$ ($I_{AA}$) and with respect to peripheral events
  ($I_{CP}$). Significant suppression is observed on the side opposing
  the trigger, while a moderate enhancement is measured on the near
  side.
\end{abstract}

The distribution of angles \df and \de between charged hadron trigger
particles and their associated partners provides valuable information
on a variety of physical processes in heavy ion collisions. The
correlation function, $\CF \equiv N^{AB}_{mixed} / N^{AB}_{same} \times
(dN^{AB}_{same}/d\Delta\phi d\Delta\eta ) /
(dN^{AB}_{mixed}/d\Delta\phi d\Delta\eta)$, describes the shape. The
per-trigger conditional pair yield is given by $Y \equiv
(N^{AB}_{same} / N^A). \CF$ (although not described here, acceptance
and efficiency correction factors are also included in practice).
\begin{figure}[hbt!] \centering
  \includegraphics[width=0.32\textwidth]{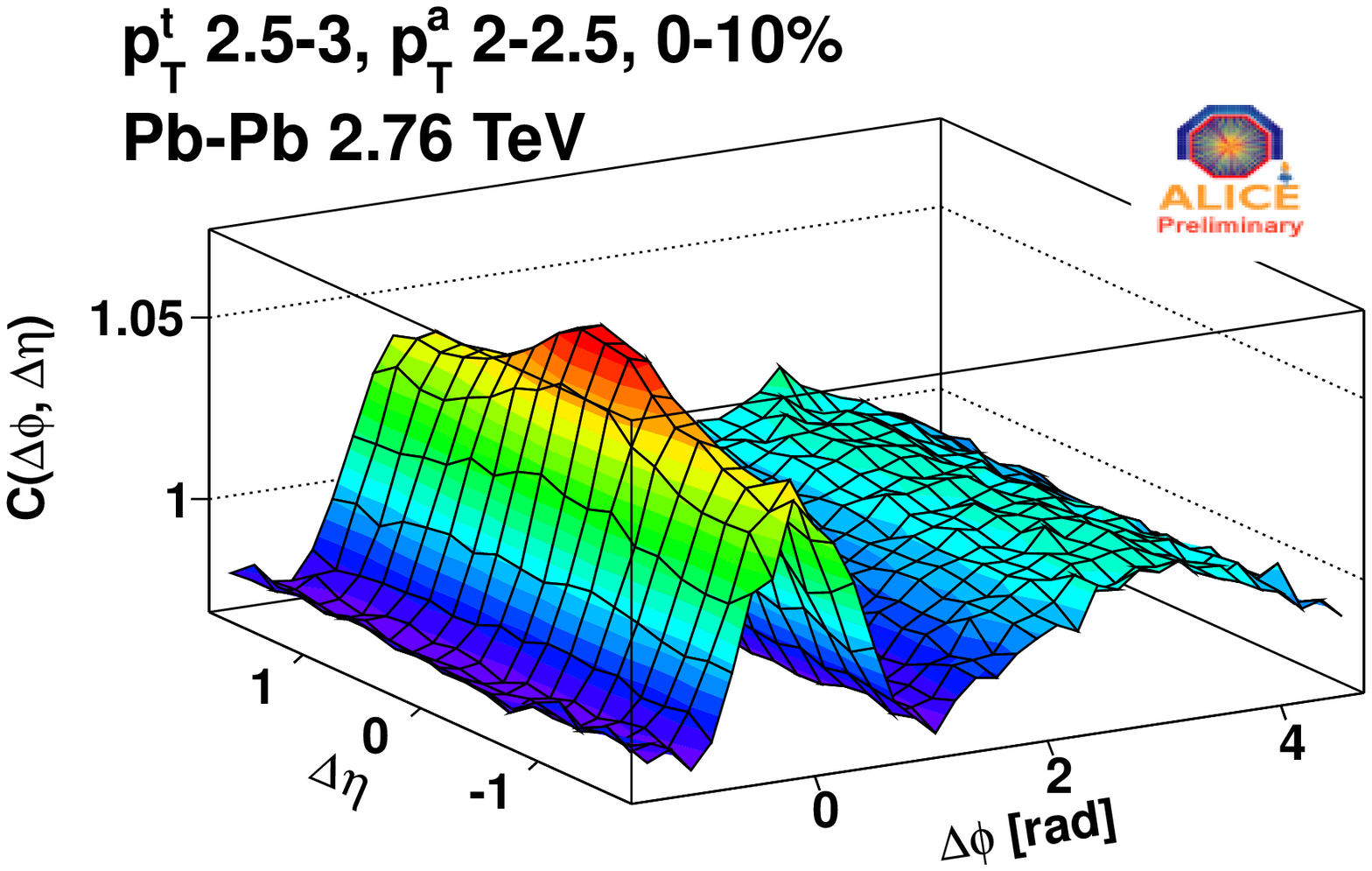}
  \includegraphics[width=0.32\textwidth]{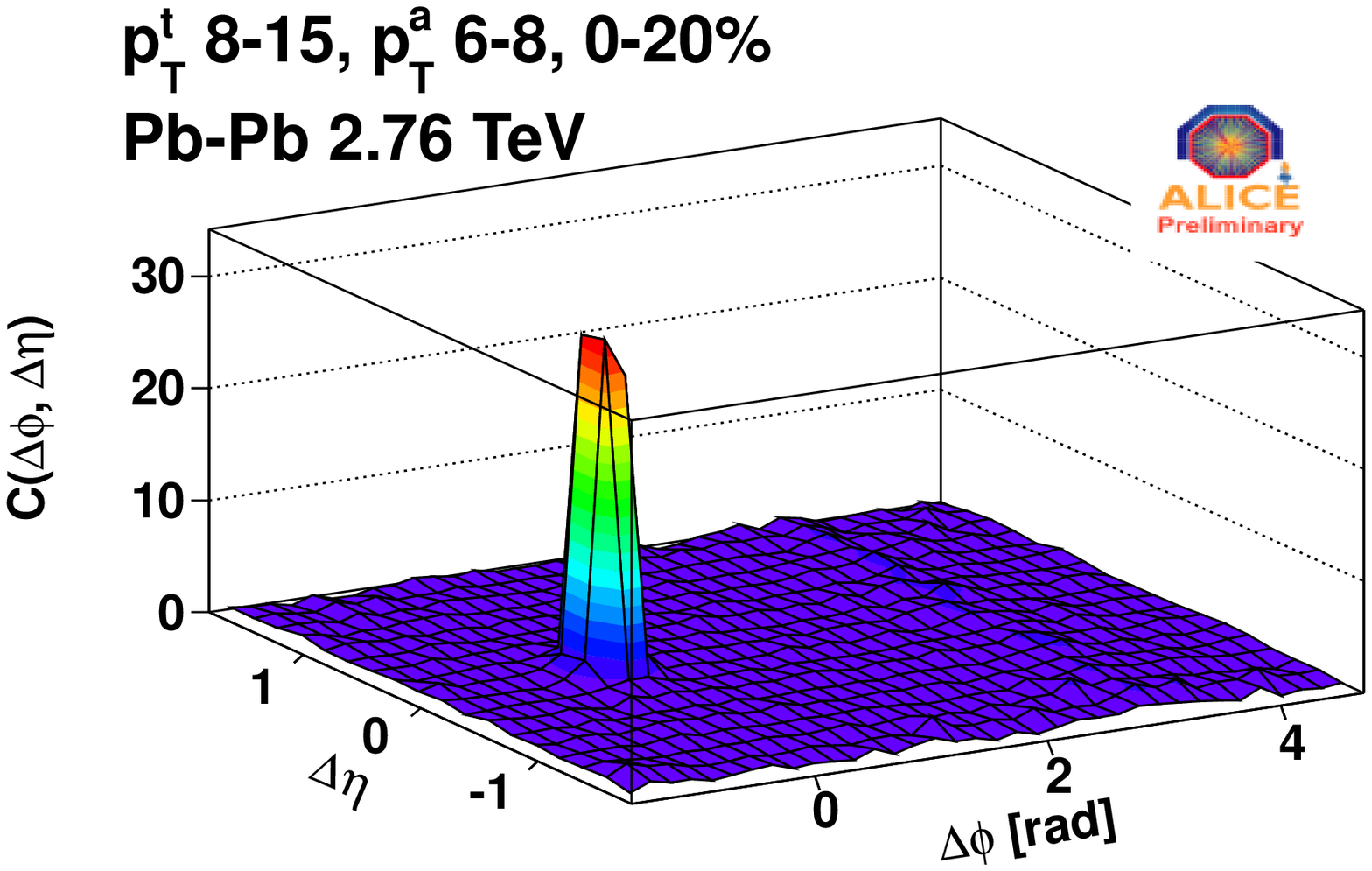}
  \includegraphics[width=0.32\textwidth]{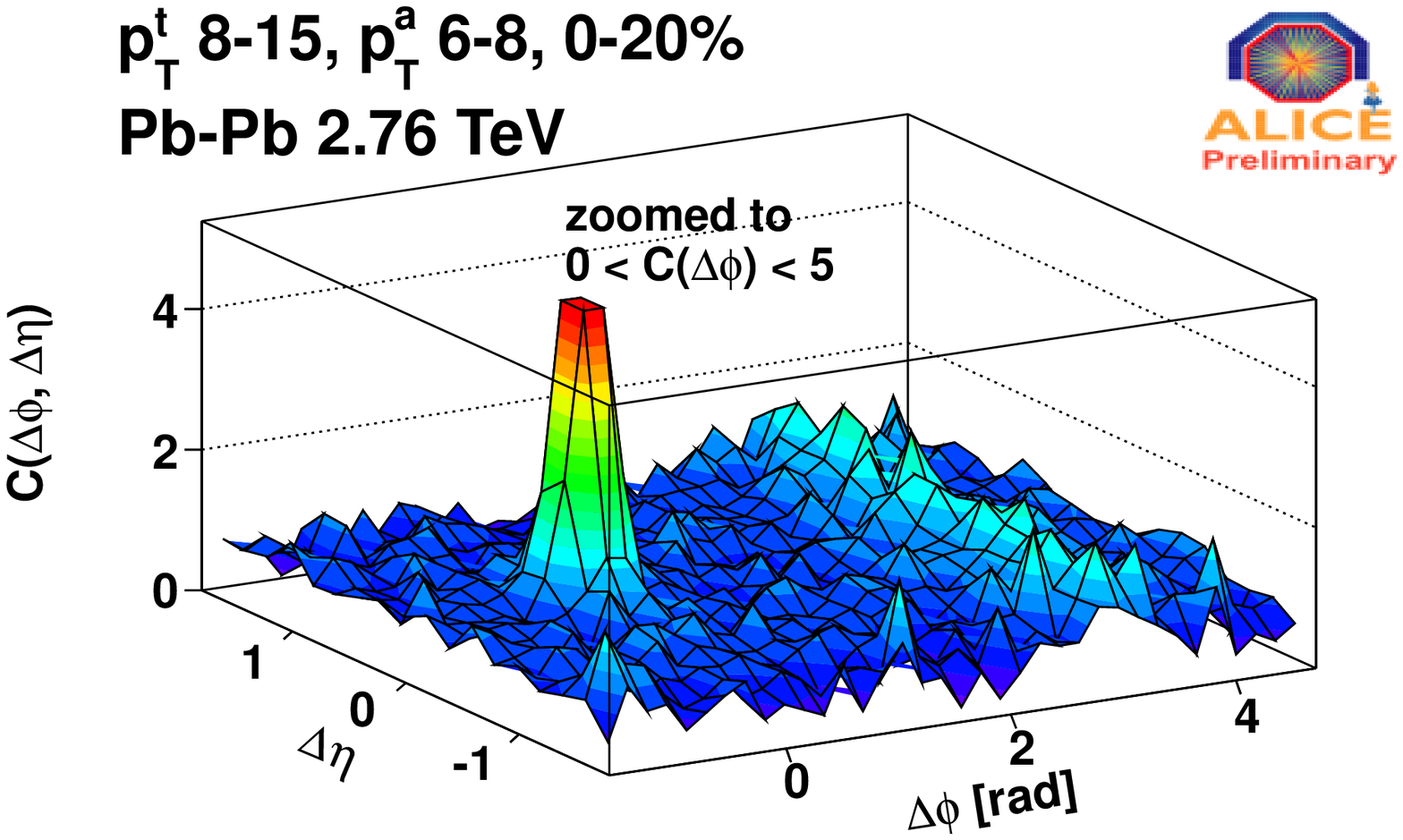}
  \caption[fig1]{\label{fig:1}$\CF$ at intermediate $\pt$ (left)
  and at higher $\pt$ (center, vertical axis zoomed at right).}
\end{figure}
At $\snn$ = 2.76 TeV, Pb--Pb events exhibit a rich evolution with
\pt~of the trigger and associated particles ($\ptt,\, \pta$), as
indicated in figure~\ref{fig:1}.  At a few GeV and higher, a peak at
$(\df, \de) \approx (0, 0)$ mainly from jet fragmentation grows in
correlation strength with \ptt and \pta, dominating the distribution
at high \pt. The away side ($\df \approx \pi$) is reduced since the
recoil jet suffers a swing in pseudorapidity due to differences
between momenta of the incoming hard-scattered partons. A ridge
feature dominates the near side at 2-3 GeV, but becomes
indistinguishable from the background at higher \pt.
\begin{figure}[hbt!] \centering
  \includegraphics[width=0.3\textwidth]{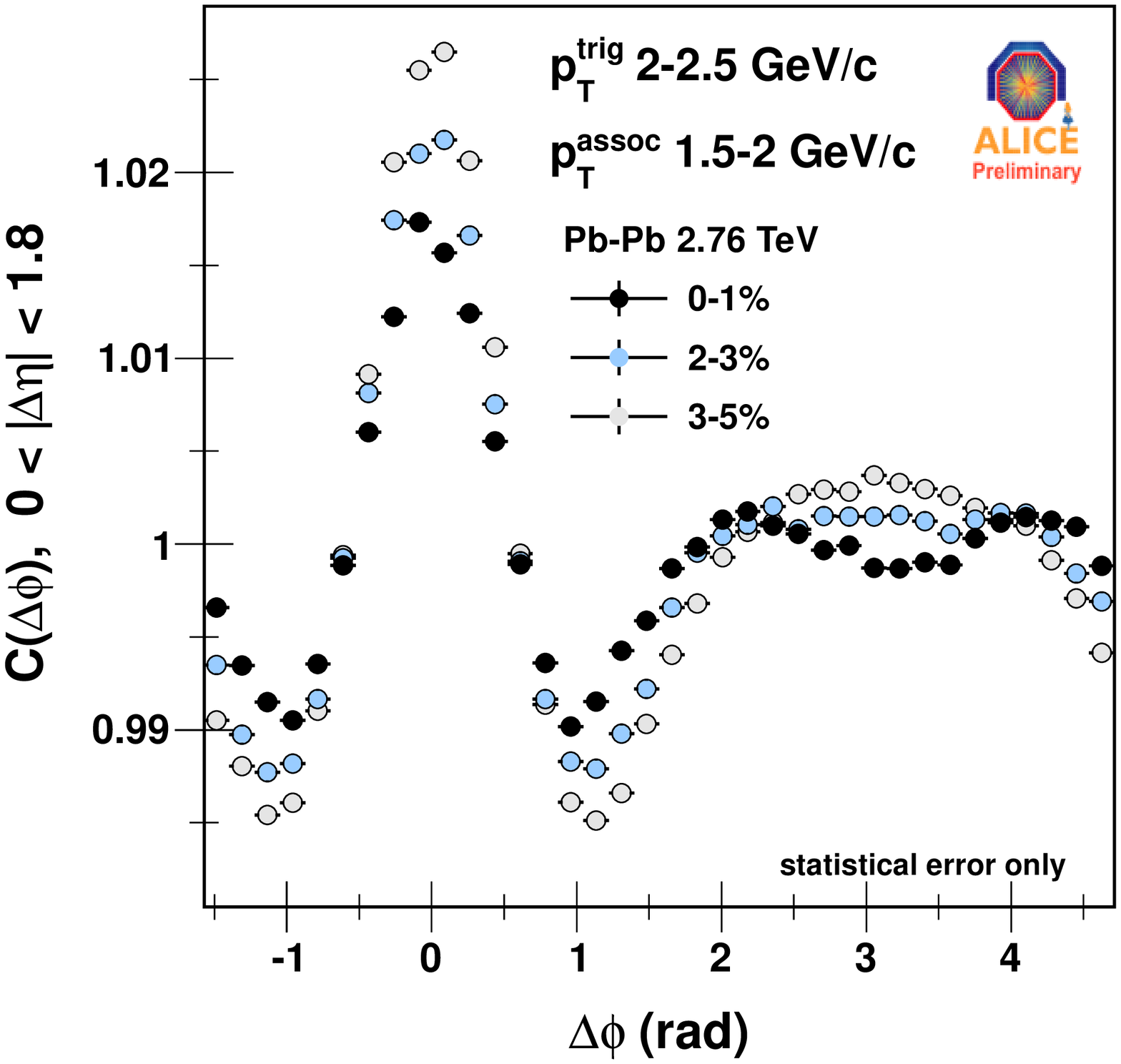}
  \includegraphics[width=0.3\textwidth]{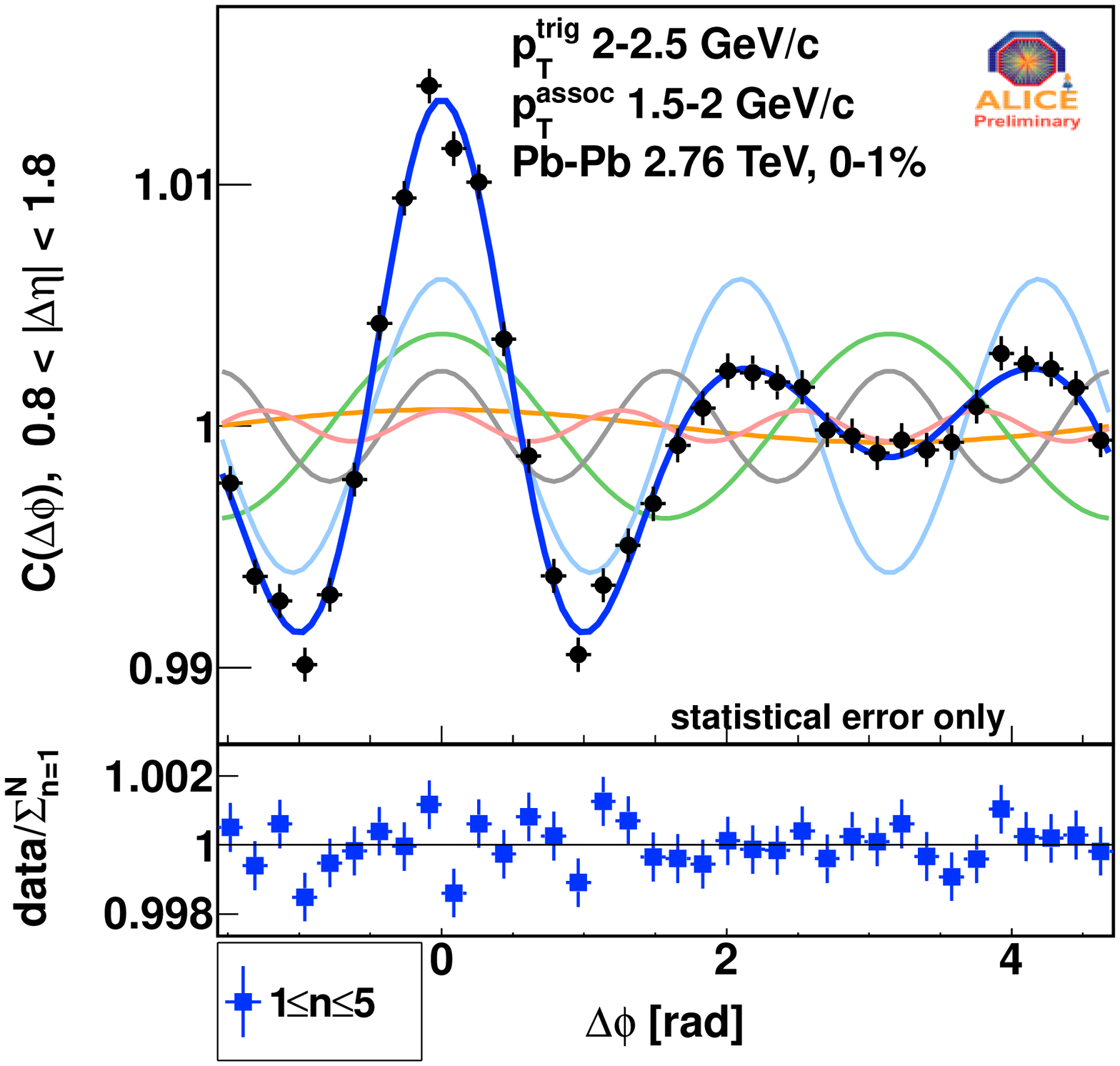}
  \includegraphics[width=0.3\textwidth]{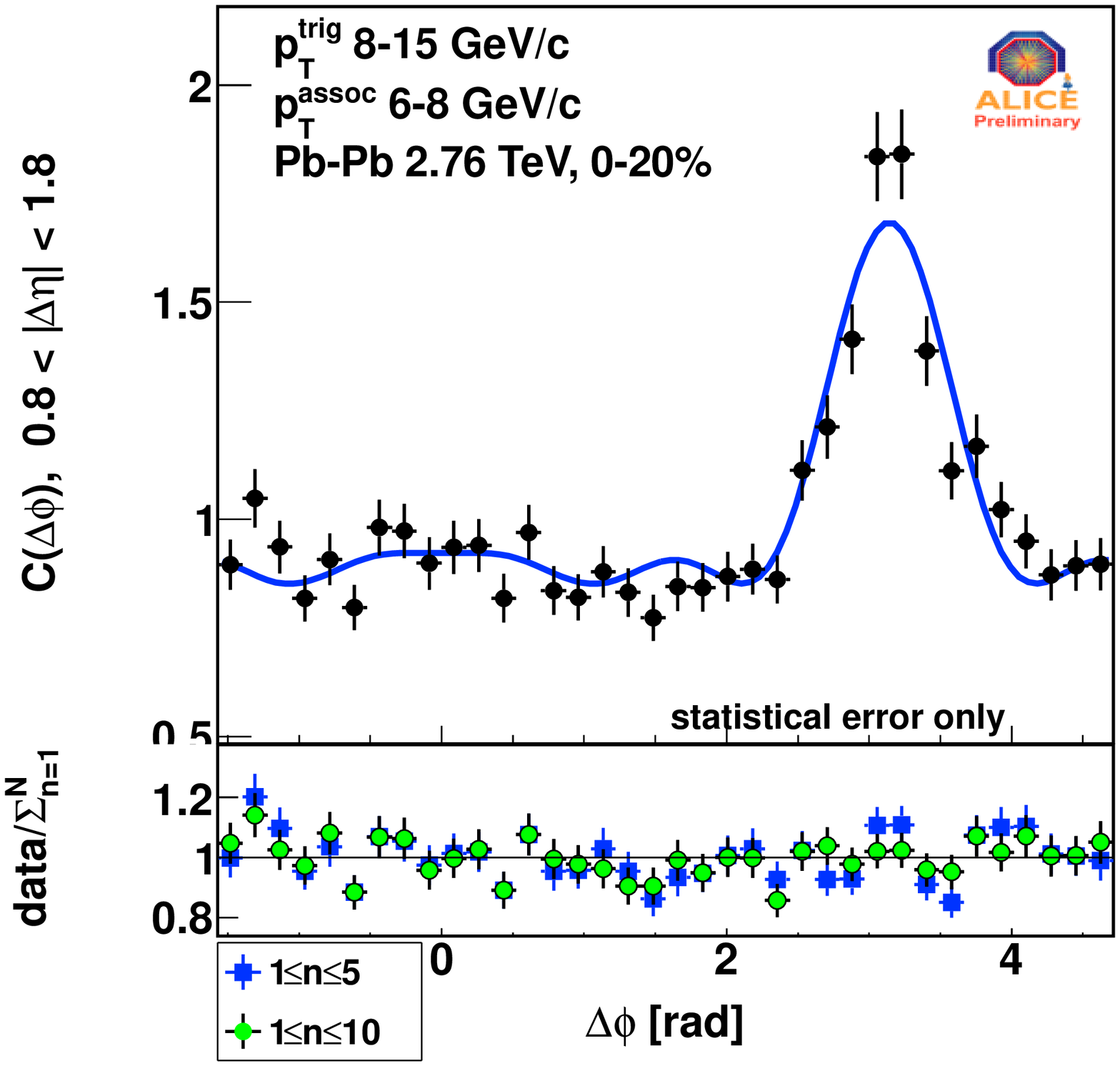}
  \caption[fig2]{\label{fig:2}Azimuthal projections of \CF for
    $0<|\de|<1.8$ (left) and $0.8<|\de|<1.8$ (center, right).}
\end{figure}
A double-peaked away side structure is visible in azimuthal
projections of the most central collisions (figure~\ref{fig:2},
left). Excluding the near side jet by requiring $|\de| > 0.8$, the near
side ridge is prominent and the doubly-peaked away side persists
(figure~\ref{fig:2}, center). The latter becomes a single narrow peak
at higher \pt, which is extended in \de and has a shape consistent
with jet fragmentation (figure~\ref{fig:2}, right).  The question
naturally emerges as to the relative contributions of the correlation
sources as a function of momentum. To this end, we study the Fourier
components of $C(\df)$: $\vnd \equiv \langle \cos n\df \rangle =
\sum_i^{N} w_i \cos n\df / \sum _i^{N} w_i$, where $w_i$ is the
content of \df bin $i$; see for example figure~\ref{fig:2}
(center). In central events, including $n>5$ gives only a marginally
improved match, particularly below $\sim 3$-$4$ GeV. At \ptt (\pta)
above 8 (6) GeV, where the jets are more collimated, higher terms may
be less negligible. However, Fourier-decomposing a narrow Gaussian
recoil jet peak, although mathematically well-defined, is less
intuitive as a descriptive mechanism compared to lower \pt~cases,
where the first few terms each have distinct physical interpretations.
\begin{figure}[hbt!] \centering
  \includegraphics[width=0.48\textwidth]{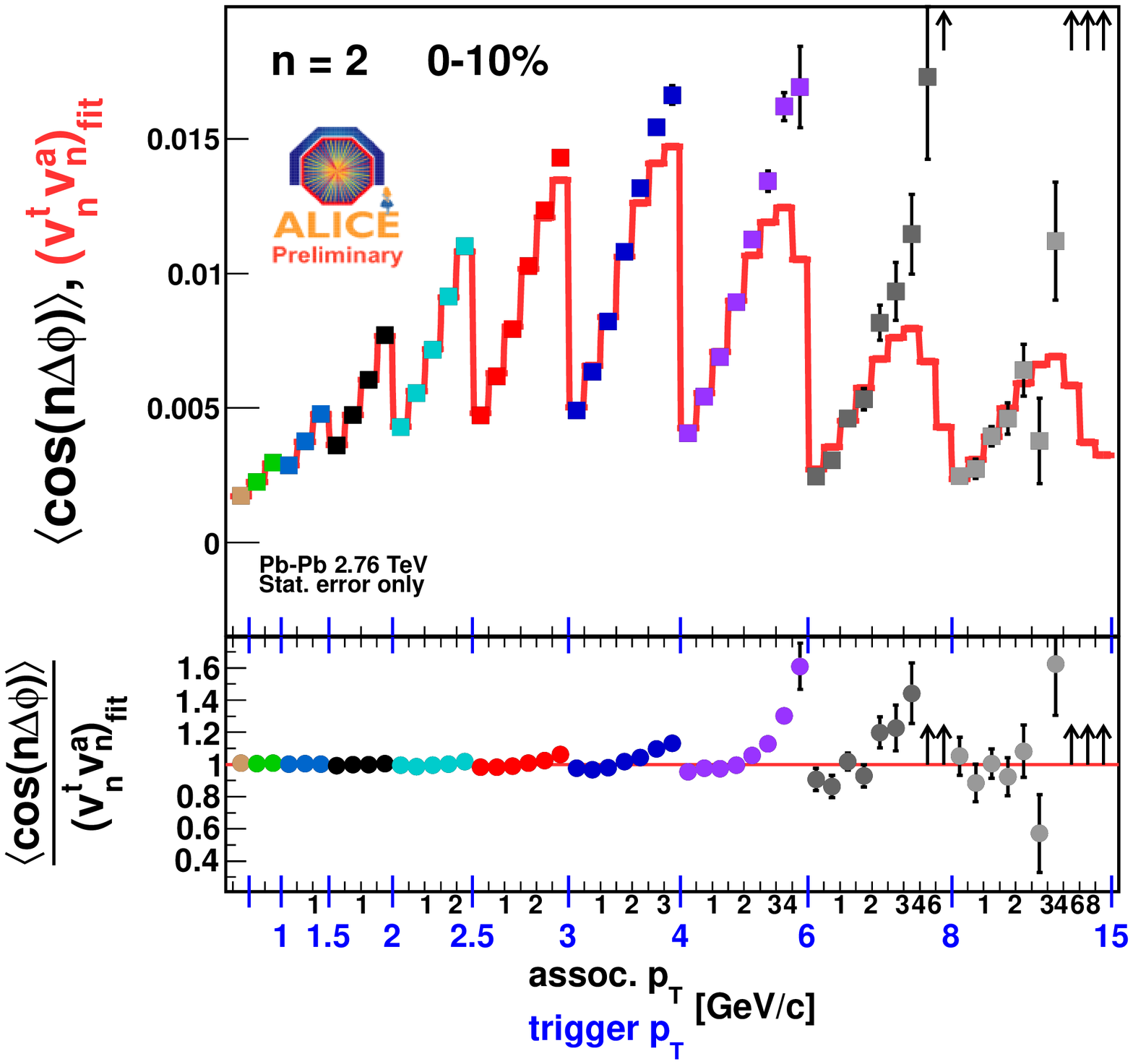}
  \includegraphics[width=0.48\textwidth]{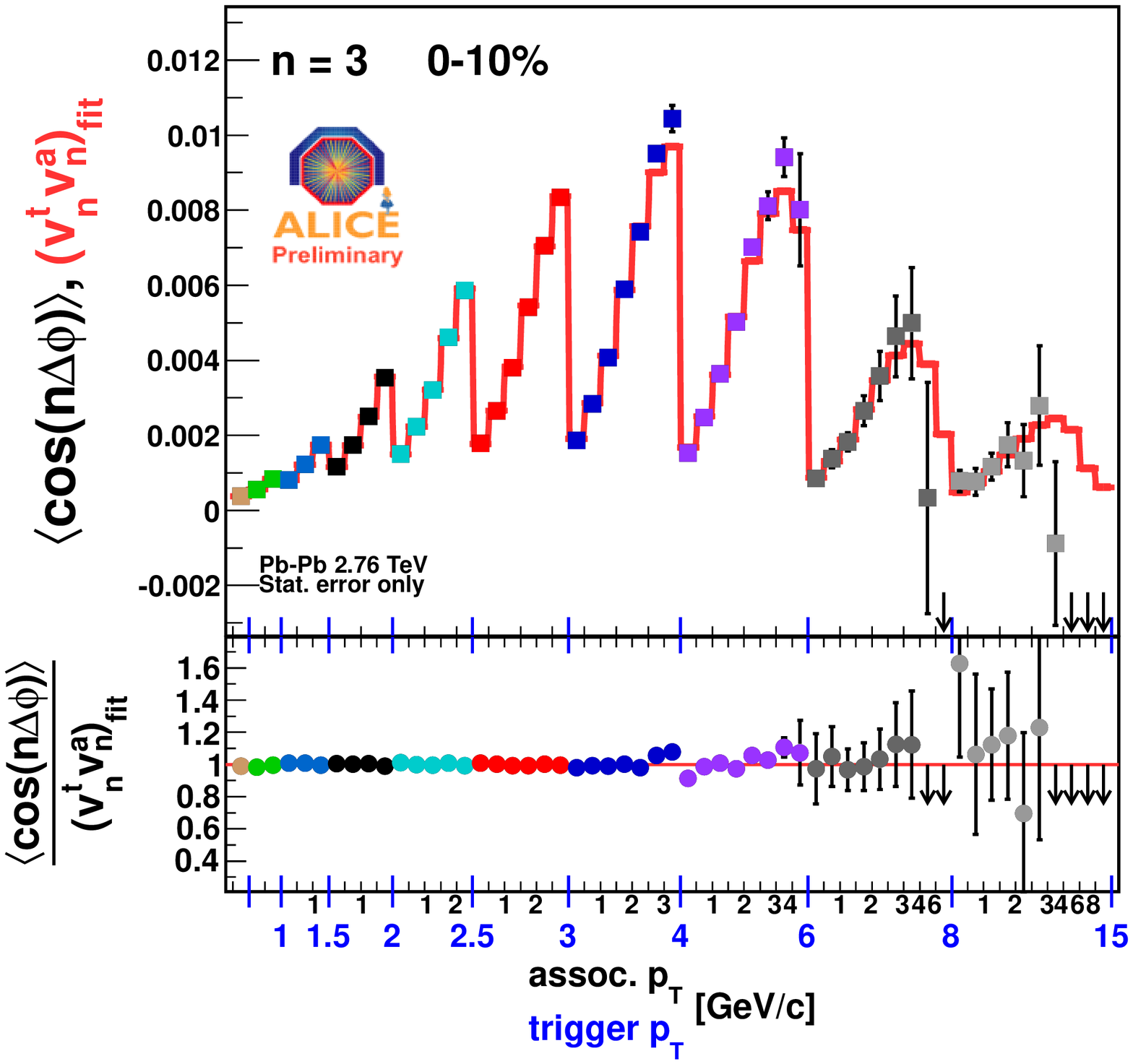}
  \caption[fig3]{\label{fig:3}Global fits to $V_{2\Delta}$ and
    $V_{3\Delta}$ in 0-10\% Pb--Pb colllisions for $|\de| > 0.8$.}
\end{figure}

If two-particle correlations arise from an event-wide collective
response to initial state anisotropy, then all particles are
``pushed'' by the same global mechanism, and should participate in a
pattern of anisotropy described by a single $v_n(\pt)$ curve. Thus,
pair correlations would be a simple factorizable product of the
single-particle bulk anisotropies~\cite{luzumPLB}:
\begin{equation}\label{eq:fac}
\vnd(\ptt, \pta) = v_n(\ptt) v_n(\pta) 
\end{equation}
In contrast, anisotropy from particles correlated only through
localized effects such as resonance decay or jet fragmentation (plus
medium response) would not be expected to factorize, since such
mechanisms have weaker direct dependence on global symmetry properties
(although effects such as pathlength-dependent energy loss can play
some role).

The factorization ansatz of eq.~\ref{eq:fac} is a testable relation,
opening the possibility to distinguish between collective and local
correlation contributions as a function of centrality and \pt. For
each $n$, the set of all $\vnd(\ptt, \pta)$ points was fit
simultaneously with the right-hand side of eq.~\ref{eq:fac} for a
variety of centrality categories. Examples are shown in
figure~\ref{fig:3}.
\begin{figure}[hbt!] \centering
  \includegraphics[width=0.9\textwidth]{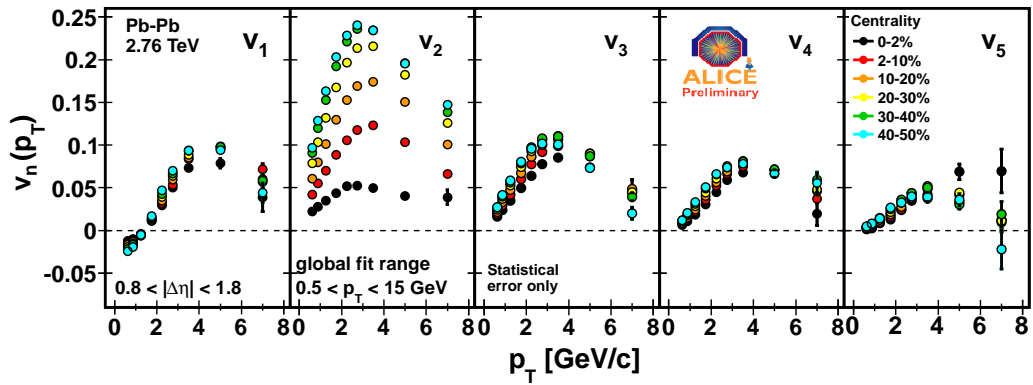}
  \caption[fig4]{\label{fig:4}$v_n$ coefficients for $n\leq 5$ produced from the
    global fit.}
\end{figure}
The measured \vnd points and the global fit generally agree for $n>1$
for $\ptt,\, \pta < 4$ GeV in central events. For all $n$, the global
factorization products tend to increasingly deviate from the data with
rising \pt~and centrality, reflecting growing nonflow contributions to
the anisotropy. The factorization for $v_1$ does not hold as well as
the others at any centrality or \pt. A dipole contribution from
\pt~conservation may be partially responsible. Corrections for this
have been proposed~\cite{luzumPRL}, but not applied here; its
inclusion is an area of current study.

To the extent that collectivity is validated by the global fit, the
resulting parameters take a physical interpretation as $v_n(\pt)$
(figure~\ref{fig:4}). The global fit $v_n$ values agree with those
obtained by more standard ``flow'' analyses such as
$v_n\{SP\}$~\cite{alicevn} for $2\leq n \leq 5$.

\begin{figure}[hbt!] \centering
  \includegraphics[width=0.8\textwidth]{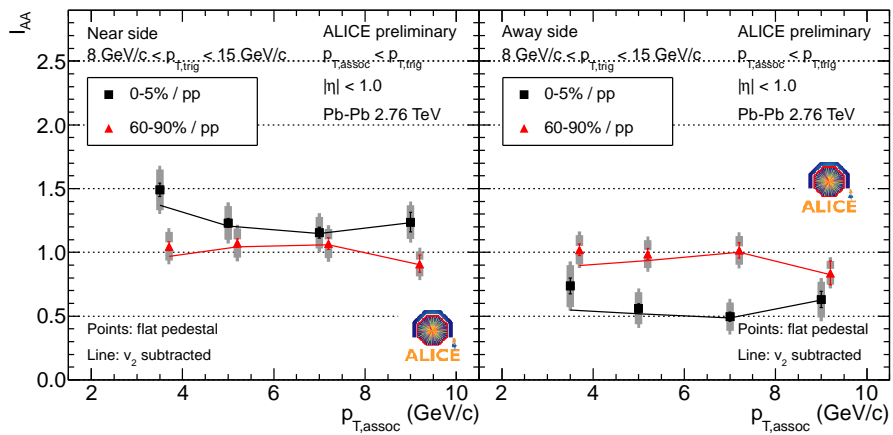}
  \caption[fig6]{\label{fig:6}Near (left) and away-side (right)
    \iaa for 0-5\% and 60-90\% centralities, where $8<\ptt<15$ GeV.}
\end{figure}

The modification of near- and away-side conditional yields from
observables such as $\iaa \equiv Y_{Pb-Pb} / Y_{pp}$ or $\icp \equiv
Y_{\mathrm{central}} / Y_{\mathrm{periph}}$ provide important
constraints on energy loss models, particularly as a complement to
\raa. At \ptt (\pta) $>$ 8(4) GeV, the correlations are dominated by
jets and the flow correlations are far smaller in comparison. In this
regime, the jet yield can be extracted with minimal biases from
assumptions about the shape and background level. The result of \iaa
at 2.76 TeV is shown in figure~\ref{fig:6}.
The small difference between using a uniform background shape (shown
as points) and a $v_2$-modulated background (solid lines) shows that
background shape assumptions do not lead to large uncertainties. We
find an enhancement of $\sim$20\% in central Pb--Pb vs. pp on the near
side, and suppression leading to \iaa $\approx$ 0.5-0.6 on the away
side. \icp was measured and also leads to the same
conclusions. Further discussion of the analysis and interpretation of
these results, as well as the \icp result, can be found
in~\cite{JFGO}.

In this presentation, the shape evolution of triggered two-particle
correlations was quantified via Fourier analysis. We conclude that the
correlation features of rapidity-separated pairs with momenta below
about 4 GeV, such as the ridge and the double-hump structure, are most
consistent with expectations from collective (i.e. hydrodynamic)
response to anisotropic initial conditions, while at higher momenta,
the breaking of \vnd factorization signals the onset of local, rather
than global, nonflow correlations. In addition, the near-side
enhancement and away-side suppression suggest the presence of a dense
medium causing a loss of energy by hard-scattered partons.

\end{document}